\documentclass[11pt,a4paper,dvips]{article}
\usepackage{a4p}
\usepackage{cite,mcite}
\usepackage{graphicx}
\usepackage{subfigure}
\usepackage{atlasphysics}
\usepackage{atlas_title,ifthen}
\usepackage{mathptmx}
\usepackage{helvet}
\usepackage{wrapfig}

\def\thedate{\today}
\newlength{\capindent}
\newlength{\capwidth}
\newlength{\figwidth}
\newcommand{\icaption}[2][!*!,!]{\hspace*{\capindent}%
  \begin{minipage}{\capwidth}
    \ifthenelse{\equal{#1}{!*!,!}}%
      {\caption{#2}}%
      {\caption[#1]{#2}}
      \vspace*{3mm}
  \end{minipage}}
\begin{document}
\begin{titlepage}
\vspace*{-6mm}
\includegraphics[width=3cm]{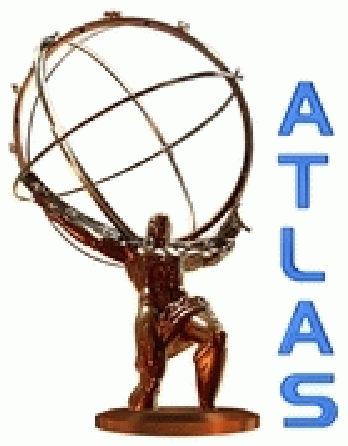} \hfill 
\begin{minipage}[b]{7cm}
\begin{center}
\mbox{\Huge \bf ATLAS NOTE} 
\end{center}
\begin{center}
\mydocversion
\end{center}
\begin{center}
\thedate
\end{center}
\end{minipage}
\hfill \includegraphics[width=3cm]{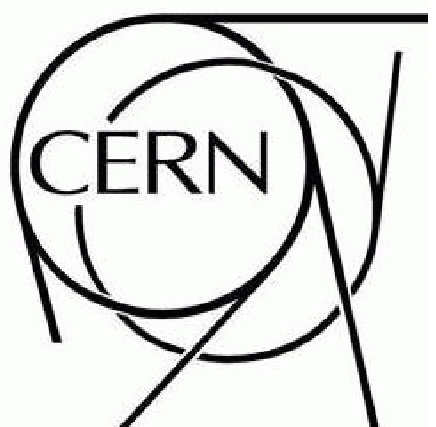}

\title{Track Based Alignment of the ATLAS Silicon Detectors with the\\
Robust Alignment Algorithm}
\author{Florian Heinemann\\University of Oxford\\f.heinemann1@physics.ox.ac.uk}
%
%
\begin{abstract}
The Robust Alignment algorithm for the ATLAS silicon
detectors is presented. It is an iterative method based on centering
residual and overlap residual distributions. Tests on simulated and
real data are discussed.  

\end{abstract}
\end{titlepage}
\tableofcontents
%
\section{Introduction}
Alignment of ever larger silicon tracking systems in modern high energy
physics experiments is
very challenging. Due to the raising numbers of degrees of freedom,
traditional alignment algorithms based on $\chi^2$ minimisation face
increasing numerical difficulties. A simple, independent and robust alignment
algorithm is crucial for a quick and reliable 
calculation of alignment constants. Furthermore, it is important to have an alternative
alignment method in order to cross-check results from $\chi^2$
minimisation algorithms, especially during early data
taking when the experimental environment is not yet well known. 

The ATLAS silicon tracking system \cite{IDTDR} consists of two subsystems: the
silicon pixel detector (PIXEL) with 1744 individual modules and the semi-conductor
tracker (SCT) which
is made of 4088 two-sided silicon strip detectors. Each of the
subsystems are divided in one barrel and two endcap detectors. 

In this paper the Robust Alignment algorithm and various tests on real
and simulated data are presented. The main advantage of this algorithm
over other track based alignment methods
is its simplicity and robustness. The Robust Alignment is limited to
the alignment of only two to three out of six degrees of freedom per module.

\section{ATLAS Coordinate Frames}
To describe the ATLAS detector various coordinate frames 
are used. In this paper the local coordinate frames defined for
each silicon module are mainly used. The origin of each Cartesian right-handed 
frame is in the centre of the module. Local Y is parallel to the beam pipe for
PIXEL barrel modules and perpendicular to the beam pipe for PIXEL endcap
modules. For SCT modules, local Y is parallel to the centre silicon
strip. The local X coordinate is always in the module plane and
perpendicular to the local Y coordinate, and is the direction in which
the silicon modules have the best spatial resolution. The local Z
coordinate is 
thus perpendicular to the module plane by construction. It always
points away from the interaction point. 


\section{Residuals and Overlap Residuals} \label{subsec:res}
The Robust Alignment algorithm is based on residual and overlap
residual measurements. Residuals are defined as the difference between
the hit and the track position in the plane of the module. The PIXEL
detector measures the hit 
position directly in both the local X and Y direction. As the SCT is
a strip detector, the local Y hit position must be calculated from
two crossing strips and the track direction. All residuals are 
defined in the local frame of the module. In general, residuals may be
biased or unbiased. Biased residuals show 
the difference between the hit 
and the track position if the track is fitted to all its hits, whereas unbiased
residuals are calculated after removing the 
respective hit and refitting the track. 

\begin{wrapfigure}{r}{.45\textwidth}
\begin{center}
\vspace{-15pt}
\includegraphics[width=.35\textwidth,angle=0]{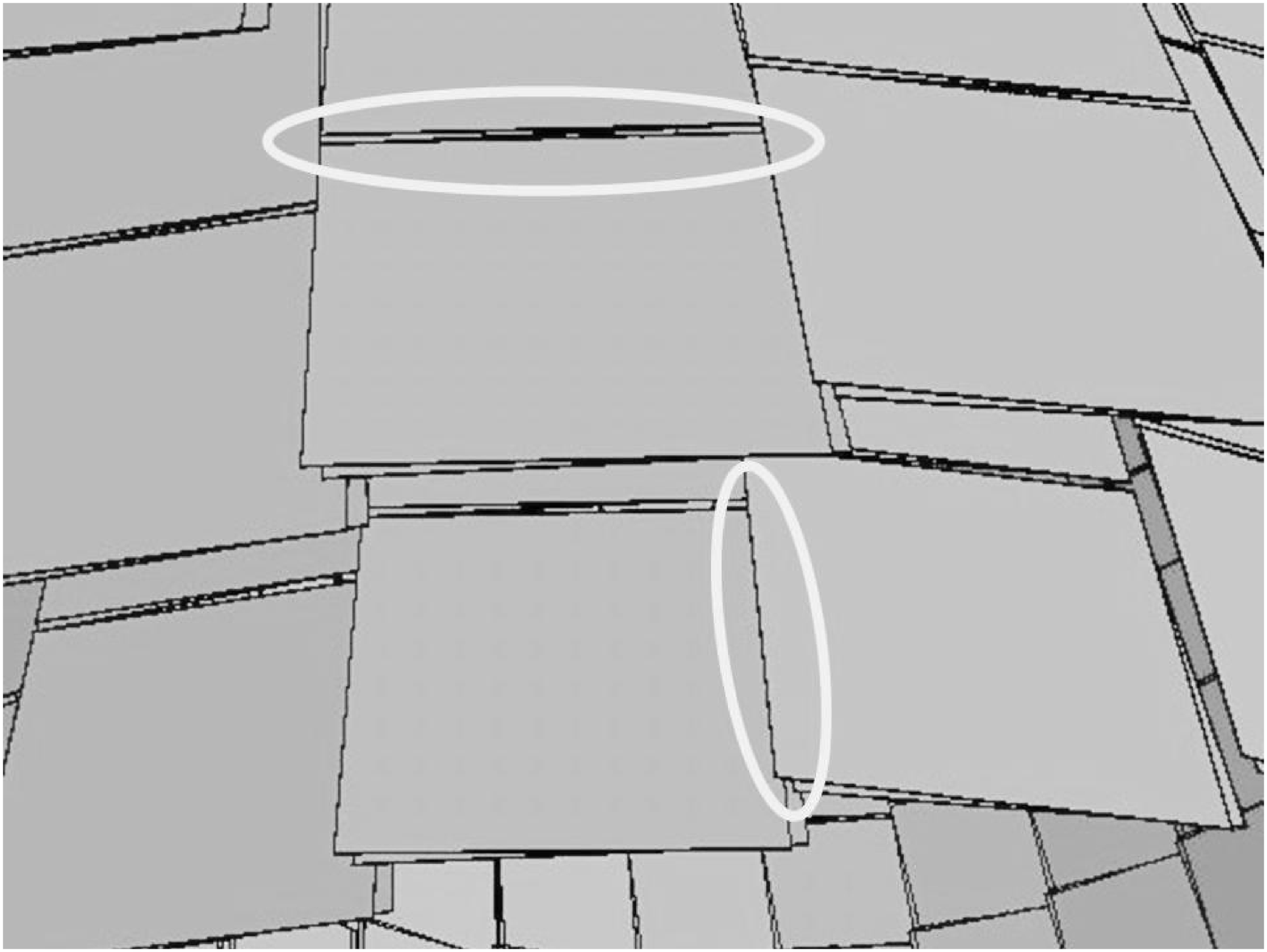}
\end{center}
\vspace{-15pt}
\caption[Overlapping silicon modules]{Overlapping silicon modules} \label{fig:overlap} 
\vspace{-20pt}
\end{wrapfigure}

Overlap residuals are defined as the difference between two residuals from
two overlapping modules (Fig. \ref{fig:overlap}).
There are two types of residuals, residuals in local X and Y, and two types of
overlaps, overlaps in the local X and Y direction. Thus, four different overlap
residuals can be constructed. The naming convention is $ovres_{AB}$
where A is the residual type and B the overlap type. Obviously, SCT
overlap residuals in local   
Y are only defined if in each module two crossing strips are hit by
a particle. 


\section{Main Principle}
\begin{figure}[ht]
\begin{center}
\includegraphics[width=.60\textwidth,angle=0]{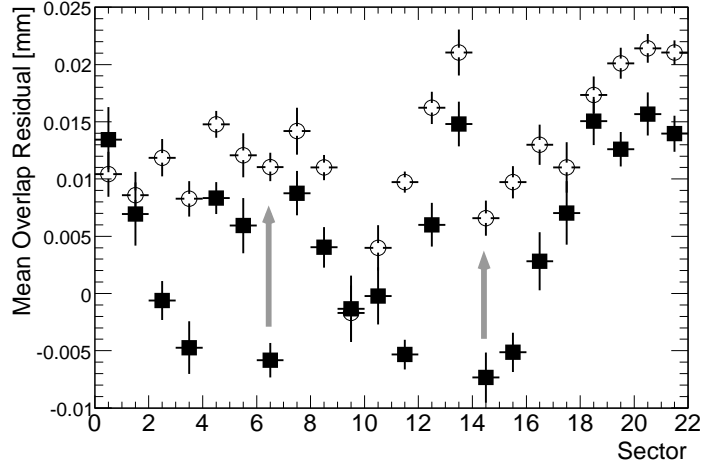}
\caption[Overlap residuals vs. radius change]{Mean overlap
residuals for all modules in a PIXEL ring where the modules are
randomly misaligned: A geometry with nominal
radius (black squares) and with a radius increased by 50 $\mum$
(white circles) are shown.\label{fig:OverlapRadius}} 
\end{center}
\end{figure}

The Robust Alignment algorithm is based on two main premises:
\begin{enumerate}
\item[$\diamond$] For a perfectly aligned detector all residual and
overlap residual distributions are centred around zero.
\item[$\diamond$] If only one module is shifted by $\delta x$ in the
direction of the residual measurement, the mean of its
unbiased residual and overlap residual distribution is equal to
$-\delta x$.
\end{enumerate}
If all modules are misaligned, correlations have to be taken into 
account. This algorithm automatically correlates modules within one
ring or stave through using overlap residuals. Dependencies on
modules which are positioned further away in the detector are taken
into account through an iterative procedure. 
Using overlap residuals it is also 
possible to take advantage of a third principle: 
\begin{enumerate}
\item[$\diamond$] The change of circumference - and thus the average
radial shift of each module - in a ring of
overlapping modules is proportional to the sum of overlap residuals in
the ring.
\end{enumerate}
If the circumference of a detector ring is at the design value, the sum
of overlap residuals vanishes even if the modules 
are generally misaligned.  However, a positive
(negative) sum means that all modules are systematically
shifted away from (towards) each other which can only be explained with an
increased (decreased) circumference.  In Fig. \ref{fig:OverlapRadius}
the change in the mean overlap residuals is shown if the radius of a
detector ring is increased by 50 $\mum$.
The measurement of deviations in the radial direction is not straight
forward for other types of algorithms, as tracks  
enter the silicon modules mostly perpendicularly. The Robust Alignment
algorithm does not measure module rotations. Furthermore, it
is assumed that the modules do not deform and that the SCT wafers
do not move with respect to each other. 

\section{Calculation of Alignment Constants}
The relative alignment shifts for each module, $a^X$, $a^Y$ and $a^Z$, are determined
from the residual and overlap residual measurements. A
shift in the mean of the residual distribution of a module and a
shift in the
mean of the overlap residual distributions is a measure of the same
quantity; the displacement of the module. For each local coordinate,
X and Y, there is one residual distribution and a maximum of two
overlap residual distributions. It is natural to combine these up to
three measurements respecting their statistical uncertainties:
\begin{equation}
  \label{eq:measurementSum}
a^{X/Y} = - \sum_{j=1}^{n} \frac{s^{X/Y}_j}{(\delta s^{X/Y}_j)^2} / \sum_{j=1}^{n}
\frac{1}{(\delta s^{X/Y}_j)^2}, \ n \leq 3.
\end{equation}
The first summand in this equation comes from the residual
measurements of each module. As these are directly correlated with the
module displacement, $s^{X/Y}_1$ is given by the mean of the module's
residual distribution:
\begin{equation}
s^{X/Y}_{1} = \overline{res}_{X/Y}.
\end{equation}
Thus, the first summand is given by:
\begin{equation}
\frac{s^{X/Y}_1}{(\delta s^{X/Y}_1)^2} = \frac{\overline{res}_{X/Y}}{\delta\overline{res}_{X/Y}^2}
\end{equation}
The statistical uncertainty on the mean
residual $\delta\overline{res}_{X/Y}$  is given by
the RMS of the residual distribution and its number of entries:
\begin{eqnarray}
\delta\overline{res}_{X/Y} = \frac{RMS_{res}}{\sqrt{n_{Hits}}},
\end{eqnarray}
If the module has a local X (local Y) overlap to a neighbouring
module the second (third) term has to be considered as
well. Overlap residuals are also 
correlated to the displacement of a 
module. Furthermore, overlap residual distributions are dependent on
displacements of the neighbouring modules. Therefore, these 
correlations have to be accounted for in the calculations of the
alignment constants. From the overlap point of view, where to
set the starting point in a ring or stave is an arbitrary choice. The
Robust Alignment 
algorithm begins with the
modules which have the smallest sector or ring number. All
other module position measurements from the 
overlap residuals are performed with respect to the neighbour with the
smallest sector or ring number:
\begin{eqnarray}
s^{X}_{2} & = & \sum_{i=0}^{i = N_{s}} \overline{ovres}_{XX}\\
s^{Y}_{3} & = & \sum_{i=0}^{i = N_{r}} \overline{ovres}_{YY}
\end{eqnarray}
Here, $N_{s}$ is the sector number and $N_{r}$ the ring number of
the module to be aligned.  
An average increase or decrease in the
overall circumference of a detector ring leads to a systematically
positive or negative bias in the sum of all overlap residuals.
If a
detector ring is complete and all overlap residuals are measured with
sufficient precision the algorithm is able to perform corrections in
the local Z direction:
\begin{equation}
a^{Z}  =  - \frac{\sum_{i=0}^{i = N_{a}}\overline{ovres}_{XX}}{2\pi}
\end{equation}
Here, $N_{a}$ is the number of all modules in the ring of the module to
be aligned. However, this bias needs to be removed from $s^{X}_{2}$ in the case of
local X local X overlap residuals:
\begin{equation}
s^{X}_2 = \sum_{i=0}^{i = N_{s}} \overline{ovres}_{XX} +
\frac{2\pi a^Z}{N_{a}}
\end{equation}
Thus, for local X measurements the second and third term in Eq. \ref{eq:measurementSum} is given by:
\begin{eqnarray}
\frac{s^{X}_{2}}{(\delta s^{X}_2)^2} & = & \frac{\sum_{i=0}^{i = N_{s}} \overline{ovres}_{XX} +
\frac{2\pi a^Z}{N_{a}}}{\delta(\sum_{i=0}^{i = N_{s}} \overline{ovres}_{XX} +
\frac{2\pi a^Z}{N_{a}})^2}\label{eq:locxlocx}\\
\frac{s^{X}_{3}}{(\delta s^{X}_3)^2} & = & \frac{\overline{ovres}_{XY}}{\delta\overline{ovres}_{XY}^2}
\end{eqnarray}
As the detector shape is a cylinder, rather than a sphere, there is no
$2\pi$-symmetry for the detector staves. Therefore, this 
correction is not applied for shifts in the local Y direction.  
The second and third term in Eq. \ref{eq:measurementSum} for the
calculation of the local Y misalignments are given by:
\begin{eqnarray}
\frac{s^{Y}_{2}}{(\delta s^{Y}_2)^2} & = & \frac{\overline{ovres}_{Y
X}}{\delta\overline{ovres}_{YX}^2}\\
\frac{s^{Y}_{3}}{(\delta s^{Y}_3)^2} & = & \frac{\sum_{i=0}^{i = N_r}
\overline{ovres}_{YY}}{\delta(\sum_{i=0}^{i = N_{r}}
\overline{ovres}_{YY})^2}\label{eq:locylocy} 
\end{eqnarray}
The alignment constants are finally given by the following equations:
\begin{eqnarray}
a^X & = & - \frac{\frac{\overline{res}_{X}}{\delta\overline{res}_{X}^2} + \frac{\sum_{i=0}^{i = N_{s}} \overline{ovres}_{XX} +
\frac{2\pi a^Z}{N_{a}}}{\delta(\sum_{i=0}^{i = N_{s}} \overline{ovres}_{XX} +
\frac{2\pi a^Z}{N_{a}})^2} +\frac{\overline{ovres}_{XY}}{\delta\overline{ovres}_{XY}^2} }{N^X}\label{eq:mainBegin}\\
a^Y & = & - \frac{
\frac{\overline{res}_{Y}}{\delta\overline{res}_Y^2}
+ \frac{\overline{ovres}_{YX}}{\delta\overline{ovres}_{YX}^2} 
+ \frac{\sum_{i=0}^{i = N_r}\overline{ovres}_{YY}}{\delta(\sum_{i=0}^{i = N_{r}}\overline{ovres}_{YY})^2}
}
{N^Y}\\
a^Z & = & - \frac{\sum_{i=0}^{i = N_{a}}\overline{ovres}_{X
X}}{2\pi}\label{eq:mainEnd}
\end{eqnarray}
The normalisations $N^X$ and $N^Y$ are defined as
\begin{eqnarray}
N^X & = & \frac{1}{\delta\overline{res}_{X}^2} +
\frac{1}{\delta(\sum_{i=0}^{i = N_{s}} \overline{ovres}_{XX} +
\frac{2\pi a^Z}{N_{a}})^2} + \frac{1}{\delta\overline{ovres}_{XY}^2} \label{eq:NX}\\
N^Y & = & \frac{1}{\delta\overline{res}_{Y}^2} + \frac{1}{\delta\overline{ovres}_{YX}^2}
+  \frac{1}{\delta(\sum_{i=0}^{i = N_{r}} \overline{ovres}_{YY})^2}\label{eq:NY}
\end{eqnarray}
The alignment
corrections are obtained in the local X, Y and Z coordinates. Therefore, from the 
residual and overlap residual point of view there is no difference
between the barrel and the 
endcaps, and the same calculations are used for both parts of
the detector.  
For large misalignments, applying damping factors to the alignment
constant calculation leads to a smoother convergence. 


\section{Uncertainties on Alignment Constants}\label{sec:uncertainties}
Statistical uncertainties on the alignment constants are determined by
the uncertainties on the mean values of the residual and overlap
residual distributions which are each given by
\begin{eqnarray}
\delta_{mean} = \frac{RMS}{\sqrt{n_{Hits}}},
\end{eqnarray}
with $n_{Hits}$ being the number of hits in that particular
distribution. Therefore, the exact uncertainty on each module shift depends on the width of the
measured distributions and on the illumination of the module. The
overall statistical uncertainties for each of the three coordinates are given by:
\begin{eqnarray}
\delta a^X & = & \frac{1}{\sqrt{N^X}} \\
\delta a^Y & = & \frac{1}{\sqrt{N^Y}} \\
\delta a^Z & = & \frac{\sqrt{\sum_{i=0}^{i = N_{a}}\delta \overline{ovres}_{XX}^2}}{2\pi}
\end{eqnarray}
$N^X$ and $N^X$ are defined in Eq. \ref{eq:NX} and \ref{eq:NY}.
The number of hits per 
module $N_{HpM}$ required to reach a certain
statistical uncertainty $\Delta$ can be estimated by
\begin{equation}
N_{HpM} = \frac{1}{\Delta^2 \times (\frac{1}{RMS_{res}^2} +
\frac{2}{G \times RMS_{ovres}^2})}\label{eq:NHpM}.
\end{equation}
Here, $RMS_{res}$ and $RMS_{ovres}$ represent the average RMS for the residual and
overlap residual distributions. The geometrical factor $G$ corrects
for two considerations. Firstly, depending on the location of the module in the detector, there
are about ten times fewer overlap residuals than residuals. Secondly,
some overlap residual measurements are correlated with others in the
same ring or stave. The uncertainties of these measurements have to be
also
taken into account. Thus, $G$ is defined as the product of the
number of correlated measurements and the ratio of residual to overlap
residual measurements. Values for $G$ are typically between 10 and 300. 
With each iteration, the residual and overlap residual distributions
generally become narrower and the tracking efficiency is improved. Thus, the
statistical uncertainties improve significantly with each
iteration. On the whole, the PIXEL detector provides a much better
intrinsic resolution compared to the SCT, leading to more precise
alignment constants. Furthermore, the number of hits per module
increase with smaller layer numbers. Hence, the innermost layers are
expected to have the best precision. 

The Robust Alignment constants have three types of additional
errors which are typical for track based alignment. First of all 
the residual measurement has a statistical uncertainty in 
itself. Both the hit position and the track position measurements have
an associated error. However, these errors are symmetric with respect to the
residual measurement. Therefore, the alignment constant calculation is not
affected by this uncertainty. 
Secondly, there are 
certain global modes which are not determined by the
alignment algorithm as the tracking is invariant under these
transformations. There are also more subtle modes which affect the
momentum but not the residual distributions. This challenge is
explored in more detail in section \ref{sec:sagittas}. 
Thirdly, as the alignment algorithm is
located at the end of the data processing chain, from 
the actual particle hitting the detector to the reconstructed track,
problems in the underlying infrastructure will certainly affect the
measurement of the module positions. 

\section{Distributed Alignment}
The Robust Alignment algorithm is suitable for distributed
alignment. Thus, the calculation of alignment constants may be performed
on many sub-samples of the input data using several computers
simultaneously. In each iteration
the reconstruction of the events in each of the sub-sample produces
a file with alignment information. In a second step these files are
merged and alignment constants calculated. This
feature is fully implemented in the ATLAS offline software
\cite{athena} and has already been used and tested successfully \cite{Heinemann}. 

\section{Sagitta Distortions}\label{sec:sagittas}
Any track based alignment algorithm is designed to measure the
relative module positions by optimising track
parameters. However, it 
is known that there are global deformations of the
detector, so called weak modes, that do not have any impact on the quality of the track
parameters. 

Tracking is invariant under global translation and
rotation. However, if the real origin $(x, y, z)$ of the tracks is
known it can be compared with the mean values, $x'$, $y'$ and $z'$, of
the vertex distribution. If $(x, y, z)$ does not match  $(x', y', z')$,
the Robust Alignment algorithm is able to perform a global transformation $T$ which is
given by
\begin{equation}
T = (x-x',y-y',z-z').
\end{equation}
Applying this correction is particularly important for a correct
measurement of the impact parameter $d_0$. 

Special care has to be taken of systematic rotations
of individual detector barrels which only affect the reconstructed
particle momentum, but not any other
parameter. Further examples of possible modes are described in
Ref. \cite{DannyThesis}. There are generally three classes of
solutions to this problem. 
Firstly, it is possible to constrain the momentum of the particle tracks used
for the alignment. Reconstructing particle tracks, especially those
originating from cosmic rays, in runs without any applied magnetic field produces
straight tracks, which is equivalent to having tracks with infinite
momentum. Thus, the momentum is naturally constrained. Other suggested
solutions include momentum constraints from invariant mass
measurements on known resonances and using the TRT to give an estimate
of the momentum. 
Secondly, tracks from cosmic rays - even with magnetic field -
correlate the upper and the lower part of the detector and thus,
constrain certain global modes. 
Thirdly, it is possible to use the Calorimeter information to detect
global sagitta distortions. With asymmetries between the E/p\footnote{E/p is the measured energy to momentum ratio.} 
distributions from
electrons and positrons it is possible to measure certain distortions independent of
the Calorimeter calibration \cite{DannyThesis}. 

The first and second solution are naturally implemented in the
Robust Alignment algorithm and have already partly been tested
successfully. For example, tests with the Combined Test 
Beam showed that the correct momentum scale can
be recovered if the momentum of a certain class of particles is
known. The third solution requires a completely independent algorithm which
translates the information from the E/p distributions into alignment constants. 


\section{Removal of Edge Channels}
\begin{figure}[h]
  \begin{center} \includegraphics[width=0.66\textwidth]{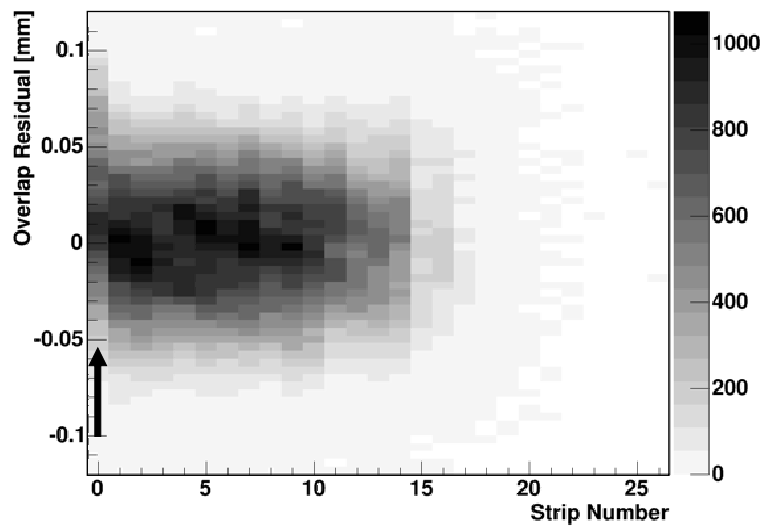}
\vspace{-4mm} 
\end{center} \caption[Edge channel bias on overlap
residuals]{Overlap residual distribution as a function of the SCT
strip number for all SCT barrels. The strip number is given for the hit
closest to the edge of the module. Only the strips which form overlaps
are presented. Channel 0 shows a positive
bias. }\label{fig:SCTEdgeChannelOvResVsStripNumber}
\vspace{-4mm} 
\end{figure}

An edge channel is defined as the strip or pixel that forms the border
of an SCT or PIXEL module. Due to the thickness of the silicon wafer, the
measured interaction point of 
the track with the silicon may lie outside of the active material if
the track is not perpendicular to the module. This biases 
the residual measurement. Furthermore, charge generated by the
particle may induce a 
signal in the neighbouring strip if the track is close to the edge of
the main strip. For edge channel strips this is, however, not possible
and again the residual measurement becomes biased \cite{CTBNIM}.
Tracks going through two overlapping modules hit edge channels more
frequently which affects the overlap residual measurement
significantly (Fig. \ref{fig:SCTEdgeChannelOvResVsStripNumber}). 
Additionally, if there is a bias in both hits of the 
overlap residual, the biases are of opposite sign. Thus, the bias of
the overlap residual is twice as big as that of the residual
measurement. Therefore, it is necessary to remove edge channels, even
though it reduces the number of overlap hits significantly\footnote{Reduction between 6\% for
SCT local X overlaps to 100\% for the central PIXEL local Y overlaps.}.


\section{Effect of Multiple Scattering on SCT Local Y Residuals}\label{sec:MCSLocalY}
Each charged particle traversing through the tracking detector is
deflected mainly by small-angle Multiple Coulomb Scattering
(MCS)\glossary{name=MCS, description=Multiple Coulomb Scattering} off
nuclei. MCS affects the tracking resolution and is most significant for low 
momentum tracks. Neglecting the uncertainty on the track fit, the
measured width $\sigma$ of the residual distribution is a function of the
intrinsic resolution $\delta_{res}$ and the average deflection due to MCS
$\delta_{MCS}$:
\begin{equation}
\sigma^2 = \delta_{res}^2 + \delta_{MCS}^2 \label{eq:resolution}
\end{equation}
Generally, the size of the average deflection $\delta_{MCS}^2$ does not
depend on the direction X or Y as the scattering in the silicon occurs
symmetrically in all directions. It is, however, not correct to assume
that the SCT detector can be considered as a simple PIXEL-like
detector in which Eq. \ref{eq:resolution} still holds. Due to the two strip detectors
geometry of the modules, their particular
placement in the different layers of the SCT and the design of the
tracking algorithm, the effect of MCS is much higher than Eq.
\ref{eq:resolution} would suggest. As the Robust Alignment algorithm
is the only ATLAS algorithm which analyses the SCT local Y residuals, this
phenomenon was unknown prior to this analysis. In the following, a
simple model is presented which correctly predicts the magnitude of the
observed increase in the local Y resolution. 

The ATLAS tracking
algorithms reconstruct tracks by minimising all 3-dimensional distances between
the track and the strips. These strips belong to modules which are
rotated from layer to layer by alternately $\pm$ 20 mrad around the
local Z axis. In order to estimate the magnitude of this effect it is
useful to make a simple, but general model \cite{Heinemann} which assumes the tracks to
be perpendicular to the module planes. One module in each of two layers
is sufficient for this purpose. Both modules are rotated with
respect to each other by an angle
$\alpha$ in the plane perpendicular to the track. Here, $\alpha$ is equal to the
stereo angle between the two wafers of each SCT module. 

Assuming that the particle undergoes MCS only between the two hits in the
respective modules, two cases are possible: a scatter in the local
X or local Y direction. In the first case,
due to geometrical effects, a scatter
$\delta_{x}$ in local X 
leads to a change in the local Y position of 
\begin{equation}
\delta_{y} = \frac{\delta_{x}}{2 \cdot \tan(\alpha)}.
\end{equation}
This term is added to the contribution from local Y scatters.
Thus, the effect of MCS on the local Y resolution gets
magnified by a factor M which can be approximated by
\begin{equation}\label{eq:MCSFactor}
M = \sqrt{\frac{1}{4 \cdot \tan^2(\alpha)} + 1}.
\end{equation}
For small angles $\alpha$ the first term is much greater
than the second. For the ATLAS SCT $\alpha$ is
40 mrad. With this model there is an increase of the MCS effect
by a factor of approx. 13. 
Simulation studies with single muon tracks
were performed to test this simple 
model. A low and a high energy sample which 
contained only muons with momenta greater than 200 MeV and 20 GeV,
respectively, were used. Using Eq.\ref{eq:resolution} and assuming
that MCS above 20 GeV is negligible, an average 
$\delta_{MCS}$ of almost 80$\mum$ for tracks above 200 MeV was
found by measuring the local X resolution $\sigma$ in both samples.
By applying then the same calculations to the local Y residual distributions
in both samples, a factor M of about 14 was measured. 
Hence, this simple model 
predicts the correct order of this effect.

The residual measurement in the local X direction is slightly
magnified due to MCS 
in the local Y direction. However, this effect is only of the order $10^{-5}$:
\begin{equation}
M = \sqrt{\frac{\tan^2(\frac{\alpha}{3})}{4} + 1}.
\end{equation}
This model suggests that the magnification of MCS effects would
neither exist if tracks were  
reconstructed by minimising 
all 3-dimensional distances between the track and the 2-dimensional space points made out of
two strips, nor would it be there if the modules had the same
orientation in each layer. In the latter case minimising 
all 3-dimensional distances between the track and the 2-dimensional space points gives the
same result as minimising the distances to all strips individually. 


\section{Combined Test Beam Alignment}\label{sec:CTB}
The ATLAS Combined Test Beam (CTB), which took place at CERN using the H8 beam
line, was a test of final prototypes of the
detector elements which had similar performances to those installed in
the full ATLAS detector. Its aim was to represent a
slice of the full ATLAS detector at $\phi=90^o$ and $\eta=0.0$
containing elements of the Inner Detector, the Calorimeter and the Muon
detectors. The data used in this study was collected in October and November 2004
after the installation of the PIXEL and SCT modules. The coordinate
system was chosen to be right handed with the x-axis in
beam direction and the y-axis vertically towards the sky. The CTB
Inner Detector consisted of three sub-detectors: PIXEL, SCT and 
TRT. The PIXEL detector was made of six modules, two in the PIXEL B
layer and two each for PIXEL 
layer 1 and 2. The active surface of each
module was $z \times y = 60.8 \times 16.4 \mm^2$. There was an overlap
of about $2 \mm$ between the two modules in each layer.
The SCT detector
consisted of four layers with two modules per layer covering each an area
of $z \times y = 120.0 \times 60.0 \mm^2$. There was a $4 \mm$ overlap between the two
modules in each layer. Although the CTB setup approximately represented
a slice of the full ATLAS detector, SCT endcap instead of barrel
modules were used.

\begin{figure}[ht]
\begin{center}	
\includegraphics[width=\textwidth,angle=0]{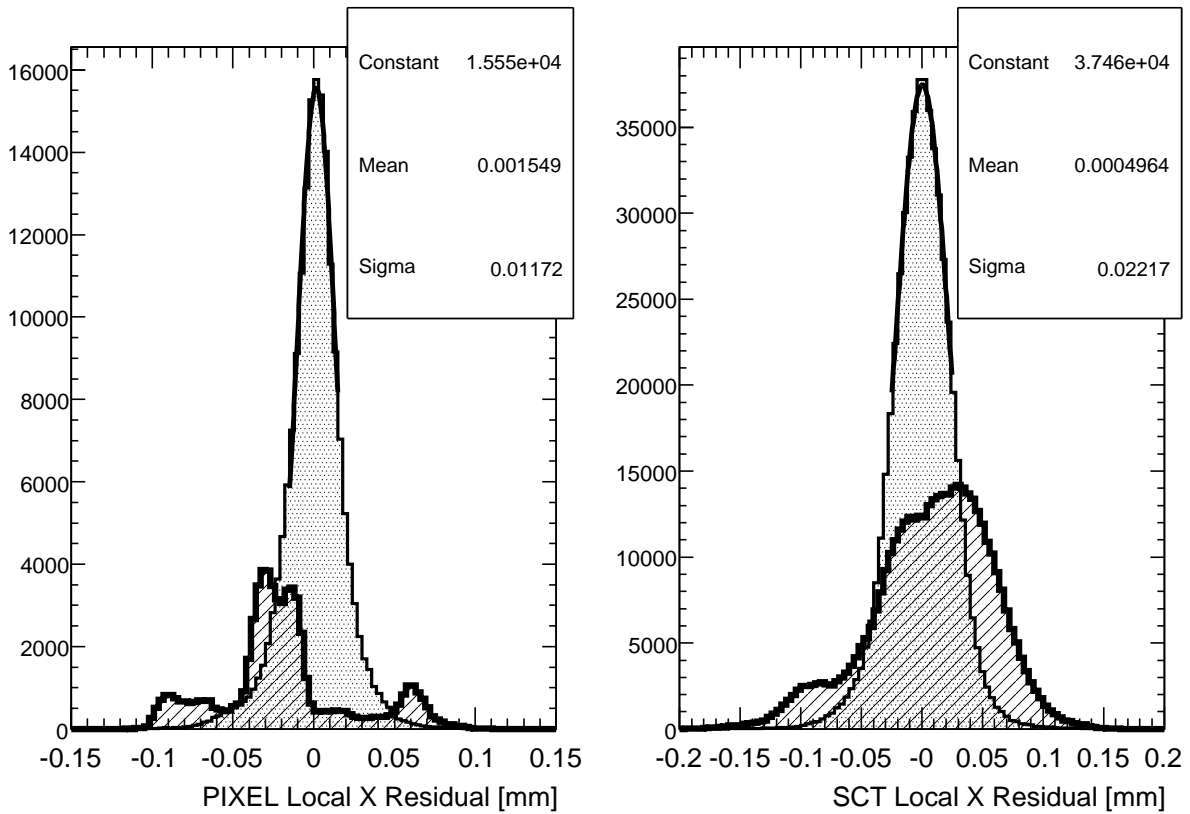}
\caption[CTB local X residuals after Robust Alignment]{The local
X residual distributions for the CTB PIXEL (left) and SCT (right)
detector are shown: Initial alignment (dashed area) and after robust 
alignment (dotted area) with run 2102355.\label{fig:CTBAlignmentRes}} 
\end{center}
\end{figure}

The Robust Alignment aligned all 14 CTB modules successfully in the
local X and Y direction \cite{CTBAlignmentNote}. The
algorithm converged quickly to a stable solution without applying any
alignment specific track selection. For the alignment presented 72 553 events with $100
\GeV$ pions unexposed to a magnetic field were used.
Although the alignment was performed with only one
run, the final alignment constants were also valid for all other
runs, with or without magnetic field.
The Robust Alignment succeeded improving the residuals significantly as can be seen in
Fig. \ref{fig:CTBAlignmentRes}. Some of the CTB modules were mounted
by hand which introduced much larger tilts than there are expected for
the full ATLAS detector. Therefore, after alignment the hit residuals
are partly still larger than it is expected from simulation.   


\section{SCT Alignment With Cosmic Ray Tracks}\label{sec:Cosmics}
The first test of the SCT barrel with cosmic ray muons was performed on
the surface inside the SR1 building at CERN. The whole setup consisted
of all four SCT barrels, the TRT and three scintillator layers which
were used to trigger cosmic ray events. Only part of the SCT
and TRT was connected to the read out system. In total 467 SCT modules
and 24 TRT modules were used to measure tracks from cosmic rays.
Due to a 15cm layer of concrete, muons with momenta smaller than
approx. 100 to \mbox{200  
MeV} did not reach the lower trigger. The collected data sample
contains only tracks from muons which did.

\begin{figure}[h]
\begin{center}	
\includegraphics[width=0.9\textwidth,angle=0]{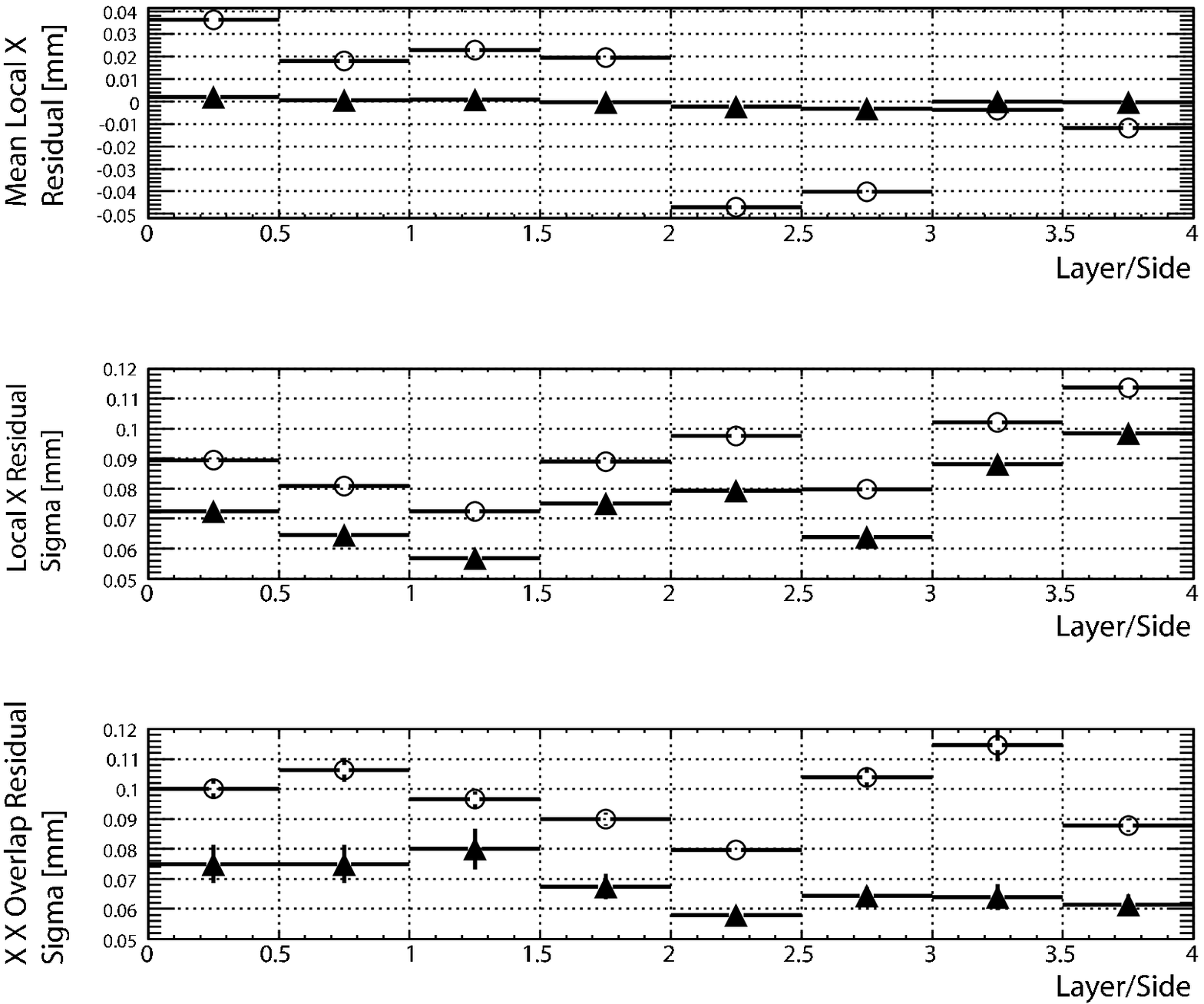}
\caption[SCT SR1 Residuals and Overlap Residuals]{Top: The mean SCT local X residuals are shown
for each layer and side before (white circles) and after Robust Alignment
(black triangles). Middle: The unbiased SCT local X residual widths are shown
for each layer and side before (white circles) and after Robust Alignment
(black triangles). Bottom: The biased SCT local X local X overlap residual widths are shown
for each layer and side before (white circles) and after Robust Alignment
(black triangles).\label{fig:SR1Residuals}} 
\end{center}
\end{figure}

The Robust Alignment algorithm successfully aligned all 467 modules of
the SR1 SCT setup in the
local X and Y direction. The
algorithm converged to a stable solution without any alignment specific
track selection. The
final alignment constants improve the residual
and overlap residual distributions
significantly (Fig. \ref{fig:SR1Residuals}). After alignment, the distributions are much
narrower and nicely centered around zero. The layer and side dependence of the
residual widths is caused by MCS and also appears in simulated data. 
For the alignment presented, 
200 000
events collected in 13 runs were used. This dataset is about half
of the available sample and 
is sufficient to produce alignment constants with a
statistical uncertainty of mostly 1 $\mum$ in \mbox{local X} and about
40 $\mum$ in local Y.
The alignment using the SR1 
data was performed with distributed alignment. Neglecting the waiting time
for the CERN batch system an overall integrated time of about 34 hours
with 10 CPUs was used for the alignment. However, already after \mbox{5 hours}
good alignment constants were produced. The final 50
iterations which lead to full convergence only improved
the resolution by a few microns.  


\section{CSC Alignment}\label{sec:FDA}
\begin{figure}[ht]
\begin{center}	
\includegraphics[width=\textwidth,angle=0]{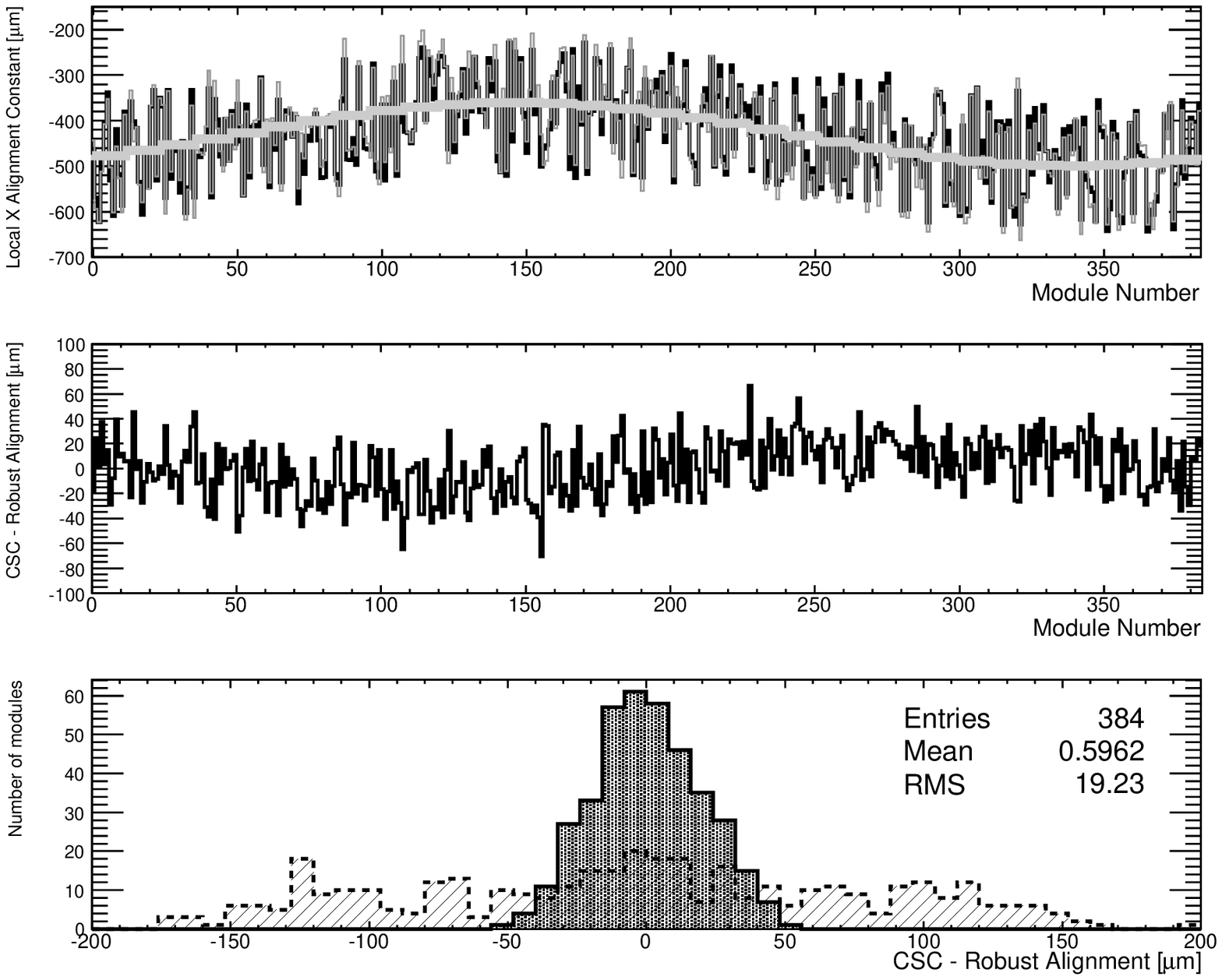}
\caption[CSC Alignment]{Alignment results for SCT barrel 0 are
shown. Top: Alignment constants for CSC geometry (black), CSC global
misalignments (thick light grey) and Robust Alignment (thin grey). Middle:
Difference between CSC geometry and Robust Alignment
constants. Bottom: Difference between CSC geometry and Robust Alignment
before (dashed) and after (dotted grey area) alignment. \label{fig:CSCAlignment}} 
\end{center}
\vspace{-5mm}
\end{figure}

In order to study scenarios where all modules are randomly
displaced and understand the effect of
systematic distortions to detector subsystems, such as the individual
SCT and PIXEL barrels
or endcap disks, data samples were
simulated with an ``as built'' geometry and a distorted magnetic
field \cite{CSCDetectorDescription}. The alignment studies (only one example is
presented) are part of the ATLAS Computing System
Commissioning (CSC) and Calibration Data Challenge (CDC). In
Fig. \ref{fig:CSCAlignment} the Robust Alignment
performance for the local X alignment of SCT barrel 0 is shown. 
This study was performed on the level of local misalignments using
events with ten muons each\footnote{The energy of these muons is
equally distributed between 2 $\GeV$
and 50 $\GeV$ with a $\eta$-range of $\pm2.7$.}. 
In the CSC geometry, this SCT
barrel has random displacements between \mbox{$\pm$ 150 $\mum$}. After Robust
Alignment most of these displacements were recovered. The remaining
difference between the actual CSC geometry and the Robust Alignment
constants is on average about 20 $\mum$. This difference is caused by
statistical effects, module rotations and remaining systematic global
distortions. With these alignment constants, residuals, track parameters
and track efficiencies are significantly improved. Studies including
global distortions showed a similarly good performance of the Robust
Alignment algorithm. However, other methods will be required to fully
remove some of the remaining sagitta distortions. 


\section{Summary and Conclusion}
A robust track based alignment method using residuals and overlap
residuals has been presented. This algorithm is numerically
very stable as no matrix inversion or minimisation is
involved. Therefore, good results were achieved in all cases without
any alignment specific track selection. It is designed to align the ATLAS silicon tracking detectors
with collision data, especially when the experimental conditions are
not yet well known. Furthermore, it is ideal for performing simple and
independent cross checks to other 
alignment algorithms. In addition to radial corrections it only measures
misalignments in the direction of the two most important degrees of freedom: the
local X and Y coordinates. These variables are crucial as they have
the most impact on the track
properties. Tests showed that, in the local X and
Y direction, the Robust Alignment is able to determine alignment
constants of similar quality 
than those from the other two ATLAS track based alignment algorithms. The Robust
Alignment algorithm has been successfully implemented into the
official ATLAS offline software and is fully operational. The method
presented is very general and can be used for the alignment of any
other silicon tracking detector with overlapping modules. 

%
%
\section{Acknowledgements}
This work has been performed within the ATLAS Collaboration and I
would like to thank collaboration members for great
support. Especially, my thanks go to the ATLAS Inner Detector
alignment group, lead by Dr. Jochen Schieck and Dr. Salva Marti, for many
fruitful discussions and to Dr. Tony Weidberg for excellent
supervision. Furthermore, I would like to thank Helen Hayward and
Kathrin Stoerig for plots and the Oxford alignment group, namely Dr. Pawel Bruckman de
Renstrom, Dr. Muge Karagoz-Unel, 
Ellie Dobson, Oleg Brandt, Dr. Stephen Gibson and Dr. Alessandro Tricoli.

%
%

%
%
\bibliographystyle{atlasstylem}
\bibliography{%
robustalignnote}

%
%
\newpage

%
%

\end{document}